\documentclass{moriond}

\usepackage{amsmath}
\usepackage{overpic}
\usepackage[numbers,sort&compress]{natbib}

\def\Etaf{\eta_{\gamma\gamma}}
\def\Etas{\eta_{\pi^0\pi^+\pi^-}}
\def\Etapf{\eta^{\prime}_{\eta_{\gamma\gamma}\pi^+\pi^-}}
\def\Etaps{\eta^{\prime}_{\gamma\rho^0}}
\def\RN#1{\uppercase\expandafter{\romannumeral#1}}

\def\be{\begin{equation}}
\def\ee{\end{equation}}
\def\bea{\begin{eqnarray}}
\def\eea{\end{eqnarray}}

\begin{document}
\vspace*{4cm}
\title{(Semi-)leptonic decays of $D$ Mesons at BESIII}

\author{ Y.~H.~Yang \\on behalf of BESIII collaboration\\}

\address{ Nanjing University, \\Nanjing 210093, People's Republic of China}

\maketitle
\abstracts{
Leptonic and semi-leptonic $D$ decays at BESIII contribute the most precise experimental measurement of $|V_{cs(d)}|$ and form factor $f_{D_{(s)}}$ in the world based on 2.93 fb$^{-1}$ and 3.19 fb$^{-1}$ data taken at  center-of-mass energies $\sqrt{s} = 3.773$ and 4.180~GeV, respectively. The largest samples at the mass threshold of the charmed hadrons $D_{(s)}$  also provide chances to extract form factors of some semi-electronic decays for the first time and together with the semi-muonic decays  we could understand lepton flavour universality better.
}

\section{Introduction}

\begin{figure}[htp]
\centering
	 \mbox{
    \begin{overpic}[width=7.0cm,angle=0]{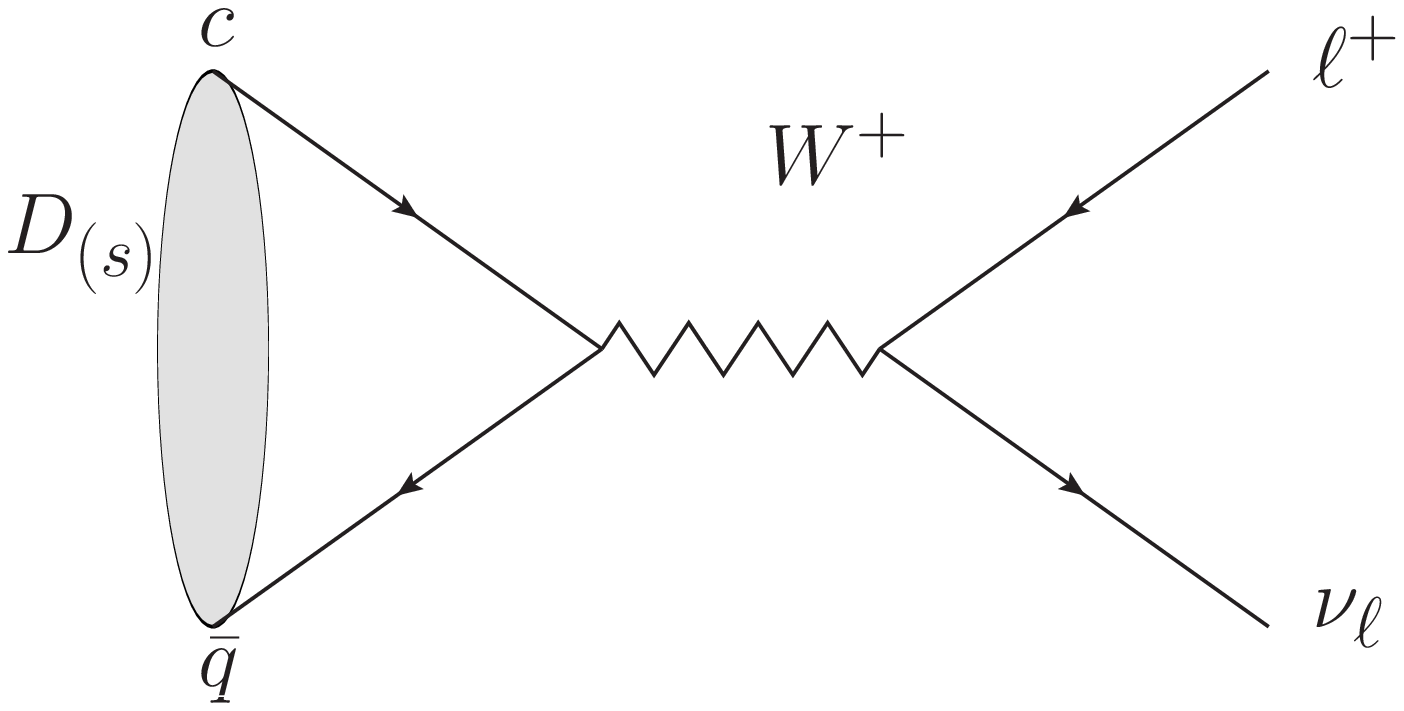}
    \end{overpic}
     \begin{overpic}[width=7.0cm,angle=0]{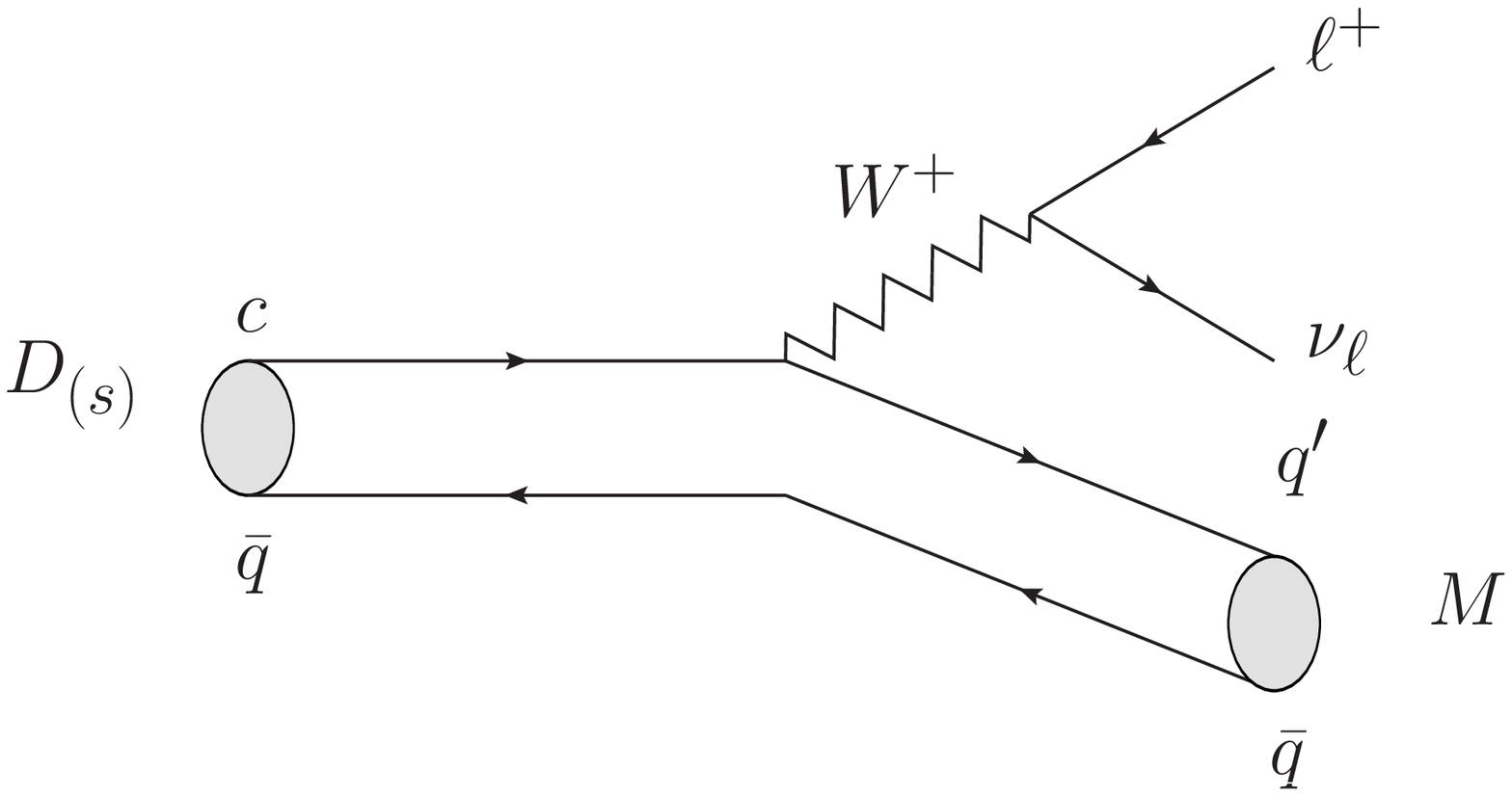}
    \end{overpic}
    }
\caption{Feynman diagrams for leptonic $D$ decays (left) and semileptonic $D$ decays to mesons (right).\label{fig:feynman}}
\end{figure}

The ground-states of charmed hadrons,  e.g., $D^{0(+)}$~\cite{Ablikim:2015gyp,Ablikim:2015qgt,Ablikim:2015ixa,Ablikim:2015mjo,Ablikim:2016sqt,Ablikim:2017twd,Ablikim:2017lks,Ablikim:2017tdj,Ablikim:2017omq,Ablikim:2018evp,Ablikim:2018lfp,Ablikim:2018frk,Ablikim:2018ffp}, $D_s^+$\cite{Ablikim:2013uvu,Ablikim:2016rqq,Ablikim:2018jun,Ablikim:2016duz,Ablikim:2018upe,Ablikim:2019rjz} and $\Lambda_c^+$~\cite{Ablikim:2015prg,Ablikim:2018woi}, can only decay weakly. Precision measurements of charm (semi-)leptonic decays provide rich information to better understand strong and weak effects as shown in Fig.~\ref{fig:feynman}. BESIII produces these charmed hadrons near their mass thresholds; this allows exclusive reconstruction of their decay products with well-determined kinematics.  For example, using $D\to \ell \nu_\ell$~($\ell=e,\mu$), we perform the most accurate measurements of $f_{D}|V_{c\bar{q}}|$, which the extraction of Cabibbo-Kabayshi-Maskawa~(CKM) matrix elements $|V_{cd(q)}|$ are essential inputs to constrain the unitarity of the CKM matrix and some first measurements of form factor $f_+^{D\to M}(0)$ by study semi-leptonic decay $D_{(s)}\to M\ell\nu_\ell$, where $M$ is a meson.  They are essential measurements to calibrate the theoretical calculation~\cite{Lubicz:2017syv,Riggio:2017zwh,Aubin:2004ej,Ball:1991bs,Na:2010uf,Bazavov:2017lyh,Bazavov:2014wgs,Boyle:2017jwu,Yang:2014sea,Bazavov:2011aa,Hwang:2009qz,Becirevic:1998ua,Na:2012iu,Aubin:2005ar,Follana:2007uv,Dimopoulos:2011gx,Chiu:2005ue,Lellouch:2000tw,Badalian:2007km} like Lattice QCD, QCD sum rule, {\it etc},  for the heavy quark decays.  The ratio of semi-muonic and -electronic decays provide an important test in the lepton flavour universality~(LFU).  

\section{Leptonic decays}
In the Standard Model, $D$ mesons decay into $\ell\nu_\ell$ via a virtual $W^+$ boson. The decay rate of the leptonic decays $D^+_{(s)}\to\ell^+\nu_\ell$ can be parameterized by the $D^+_{(s)}$ decay constant $f_{D^+_{(s)}}$ via~\cite{Silverman:1988gc}
\begin{equation}
	\Gamma(D^+_{(s)}\to\ell^+\nu_\ell) = \frac{G^2_F}{8\pi}|V_{cd(s)}|^2f^2_{D^+_{(s)}}m^2_\ell m_{D^+_{(s)}}(1-\frac{m^2_\ell}{m^2_{D^+_{(s)}}}),
\end{equation}
where $G_F$ is the Fermi coupling constant, $|V_{cs}|$ is the quark mixing matrix element, $m_\ell$ and $m_{D^+_{(s)}}$ are the lepton and $D^+$ masses, respectively. Using the measured branching fractions~(BF) of these decays, one can determine the product of $f_{D_{(s)}^+}|V_{cd(s)}|$. By taking the $f_{D_{(s)}^+}$, calculated in LQCD, or $V_{cd(s)}$, obtained from a global fit to other CKM matrix elements that assumes unitarity, the $|V_{cd(s)}|$ or $f_{D_{(s)}^+}$ can be obtained.  %In addition, the precision measurement of electron and muon leptonic decay  flavor university test is a 
	\subsection{$D^+\to\ell^+\nu_\ell$}
\begin{figure}[htp]
	\centering
	\begin{minipage}[t]{0.49\linewidth}
	\includegraphics[width=0.95\linewidth]{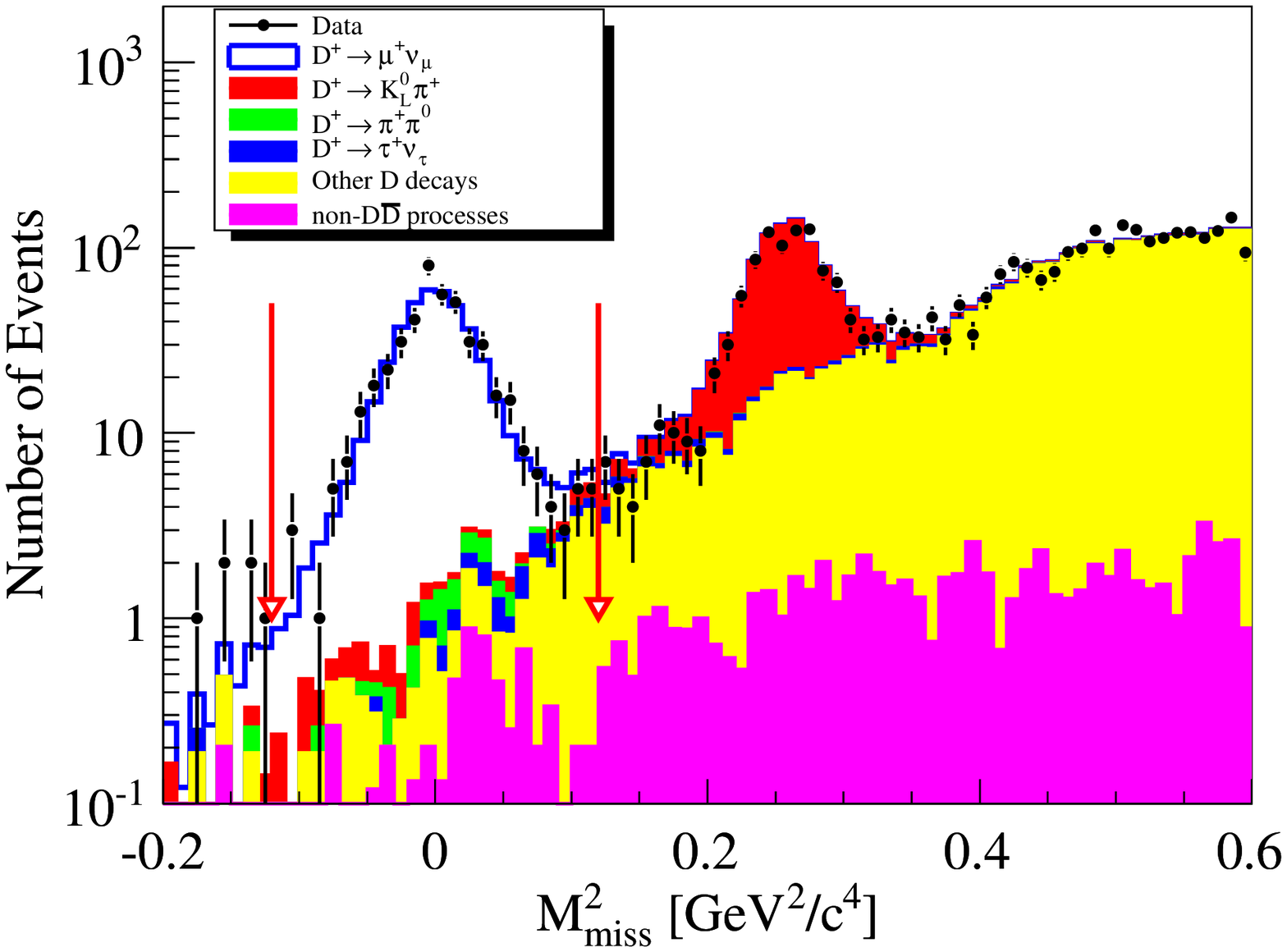}
	\caption{\label{fig:DMu}The $M^2_{\rm miss}$ distributions of the accepted candidates of $D^+\to \mu^+\nu_\mu$. Description of each background can be found on figure.}
	\end{minipage}
\begin{minipage}[t]{0.49\linewidth}
	\includegraphics[width=0.99\linewidth]{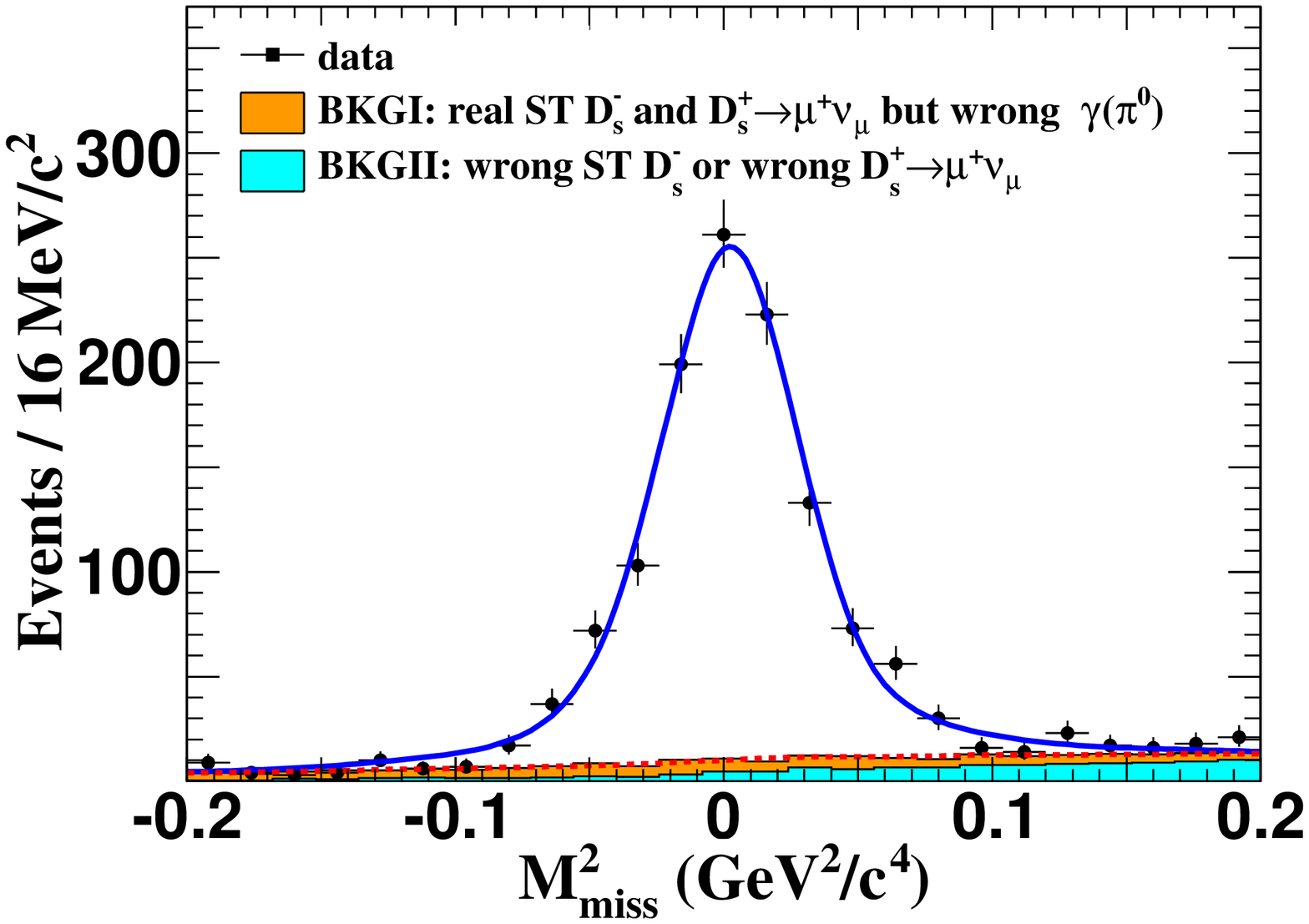}
	\caption{\label{fig:Ds_munu}Fit to the accepted $D^+_s\to\mu^+\nu_\mu$candidate events. The dots with error bars are data. The blue solid curve is the fit result. The red dotted curve is the fitted background.}
	\end{minipage}
	\end{figure}

	This analysis is based on the 2.93 fb$^{-1}$ data sample taken at the center-of-mass energy of $\sqrt{s}=3.773$ ~GeV. 
	With a total number of about $1.7\times10^6$ single tagged $D$ mesons reconstructed~($K^+\pi^-\pi^-$, $K^0_S\pi^-$, $K^0_S K^-$, $K^+K^-\pi^-$,
$K^+\pi^-\pi^-\pi^0$, $\pi^+\pi^-\pi^-$, $K^0_S\pi^-\pi^0$,
$K^+\pi^-\pi^-\pi^-\pi^+$, and $K^0_S\pi^-\pi^-\pi^+$), we obtain $409\pm21$ signals  for $D^+\to \mu^+\nu_\mu$ decay shown in Fig.~\ref{fig:DMu}. The BF of $D^+\to \mu^+\nu_\mu$ is $\mathcal{B}_{D^+\to \mu^+\nu_\mu} = [3.71 \pm 0.19 (\rm stat) \pm 0.06 (\rm sys)]\times10^{-4}$, and in conjunction with the Cabibbo-Kobayashi-maskawa matrix element $|V_{\rm cd}|$~determined from a global Standard Model fit, it implies a value for the weak decay constant $f_{D^+}=[203.2 \pm 5.3(\rm stat) \pm 1.8(\rm syst)]$~MeV~\cite{Ablikim:2013uvu}.

	BESIII also searches for the leptonic decay $D^+\to \tau^+\nu_\tau$. The preliminary result of BF is $\mathcal{B}_{D^+\to \tau^+\nu_\tau} = 1.20\pm0.24(\rm stat)\times 10^{-3}$. Combing $\mathcal{B}_{D^+\to \mu^+\nu_\mu} $, we obtain $R = \frac{\mathcal{B}_{D^+\to \tau^+\nu_\tau}}{\mathcal{B}_{D^+\to \mu^+\nu_\mu}} = 3.21\pm$0.64, which is consistent with the leptonic flavor universality in the SM prediction.
	
	\subsection{$D^+_s\to\mu^+\nu_\mu$}
	The analysis of $D^+_s\to\mu^+\nu_\mu$~\cite{Ablikim:2018jun} is based on the 3.19 fb$^{-1}$ data sample taken at $\sqrt{s}=4.178$~GeV. Using 14 ST modes, $D_s^-\to K^+K^-\pi^-$,~$K^+K^-\pi^-\pi^0$,~$K^0_SK^-$, $\Etaf\pi^-$, $\Etas\pi^-$, $\pi^+\pi^-\pi^-$, $K_S^0K^+\pi^-\pi^-$, $K_S^0K^-\pi^+\pi^-$, $\Etapf\pi^-$, $\Etaps\pi^-$,$K_S^0K_S^0\pi^-$, $K_S^0K^-\pi^0$, $K^-\pi^+\pi^-$ and $\Etaf\rho^-$, we obtain signal yield of $1135.0\pm33.1$ by fitting the $M_{\rm miss}^2$ as shown in Fig.~\ref{fig:Ds_munu}. We obtain the most precision measurement of $\mathcal{B}_{D^+_s\to\mu^+\nu_\mu}=[5.50\pm0.16(\rm stat)\pm0.15(\rm syst)]\%$ and $f_{D_s^+}=252.9\pm3.7(\rm stat)\pm3.6(\rm syst)$.

\section{Semi-leptonic decays $D\to M\ell^+\nu_\ell$}
In the SM, the weak and strong effects in SL $D$ decays can also be well separated. Their differential decay rate can be simply written as 
\begin{equation}
	\frac{d\Gamma}{dq^2} = \frac{\mathcal{B}_{D\to M\ell^+\nu_\ell}}{\tau_{D_{(s)}}} = X\frac{G_F^2}{24\pi^3}|V_{cs(d)}|^2p_{M}^3|f_+^{M}(q^2)|^2,
\end{equation} 
where $X$ is a multiplicative factor due to isospin, which equals to $1/2$ for the decay $D^+\to\pi^0e^+\nu_e$ and 1 for the other decays,
$G_F$ is the Fermi coupling constant, $p_{M}$ is the meson momentum in the $D$ rest frame, $f_+^{M}(q^2)$ is the form factor of hadronic weak current depending on the square of the transferred four-momentum $q = p_D-p_{M}$.  Based on analyzing the dynamics of SL decays, one can obtain the product of $f_+^M(0)$ and $|V_{cd(s)}|$. The form factor $f_+^M(0)|V_{cs(d)}|$ can be extracted from a fit to the measured partial decay rates in separated $q^2$ intervals. 
\subsection{$D\to \bar K\,(\pi)e^+\nu_e$}
\begin{figure*}[h!bt]
\begin{minipage}[t]{0.49\linewidth}
\centering
\includegraphics[width=\textwidth]{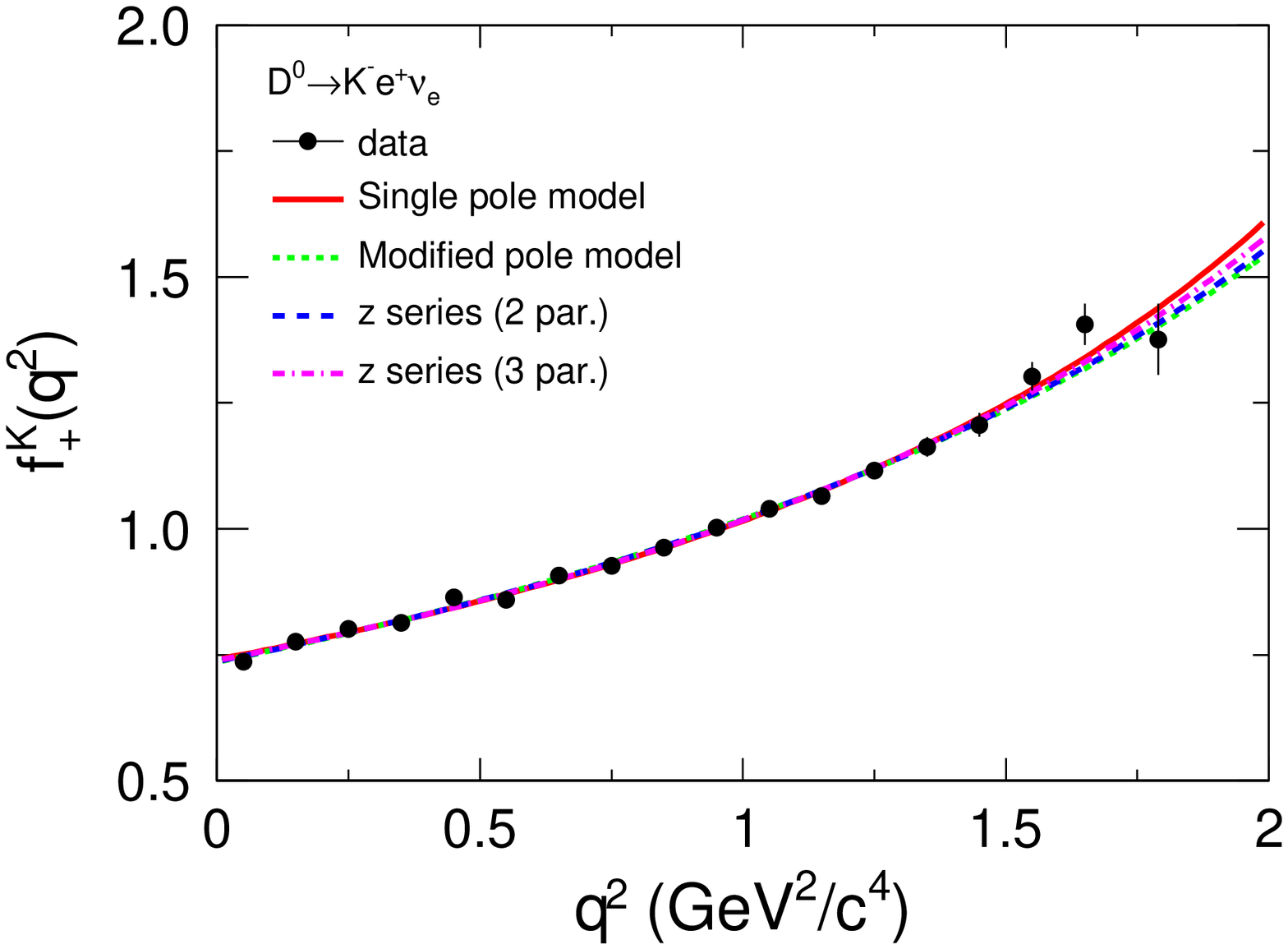}
\caption{Projection on $f_+^K(q^2)$  for $D^{0}\to K^{-}e^{+}\nu_{e}$.}
\label{projections_of_Fits_onto_ff_q2_stat_Kev}
\end{minipage}
\begin{minipage}[t]{0.49\linewidth}
\centering
\includegraphics[width=\textwidth]{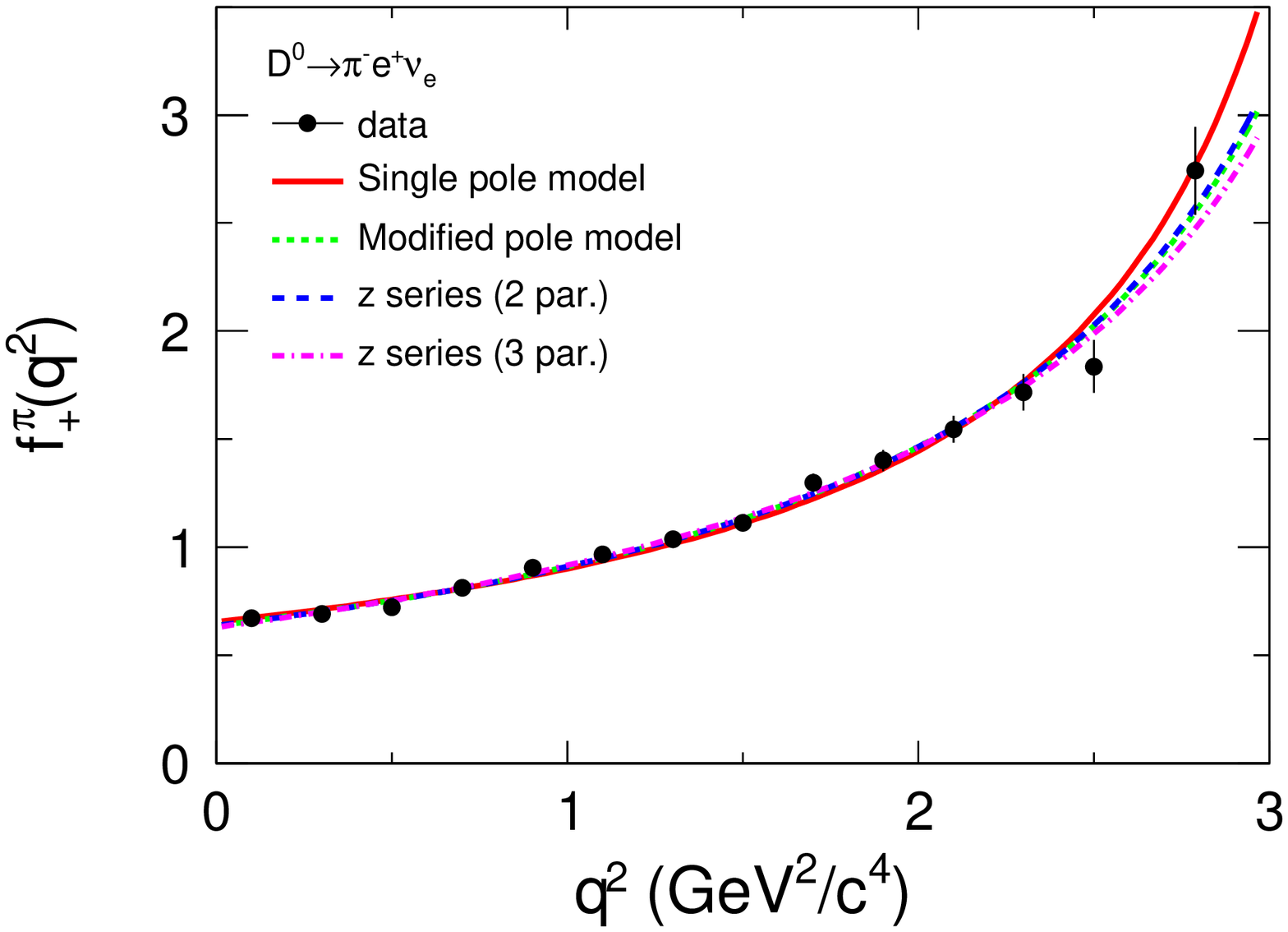}
\caption{Projection on $f_+^{\pi}(q^2)$  for $D^{0}\to \pi^{-}e^{+}\nu_{e}$.}
\label{projections_of_Fits_onto_ff_q2_stat_piev}
\end{minipage}
\end{figure*}

\begin{figure*}%[!hbp]
\centerline{
\includegraphics[width=0.45\textwidth]{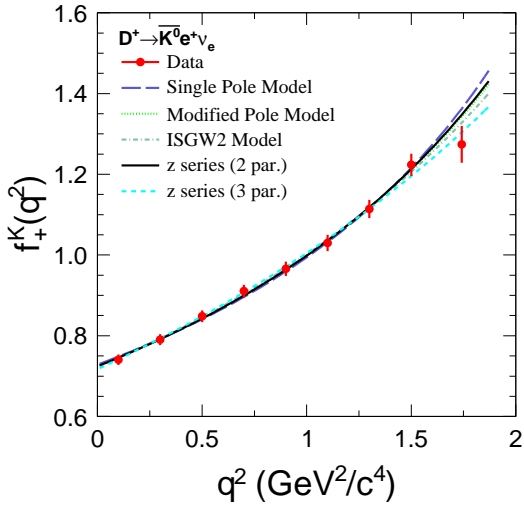}
\includegraphics[width=0.45\textwidth]{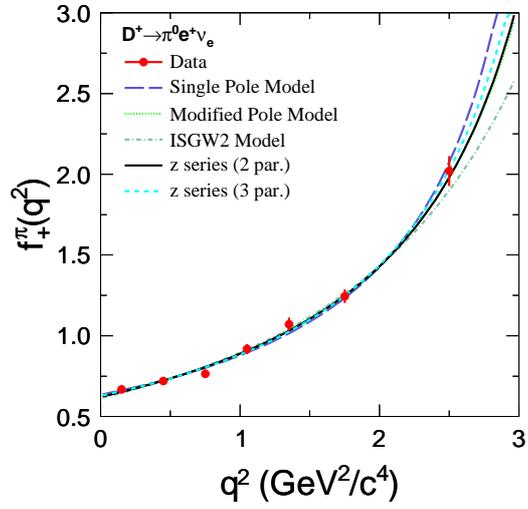}
}
\caption{
Projections on $f_+(q^2)$  for $D^+\to \bar K^0 e^+\nu_e$ (left)
and $D^+\to \pi^0 e^+\nu_e$ (right) as function of $q^2$,
where the dots with error bars show the data
and the lines give the best fits to the data
with different form factor parameterizations.
}
\label{fig:FF}
\end{figure*}

Using the same data as that of the measurement of $D^+\to\mu^+\nu_\mu$, BESIII has measured the BF of $D\to K(\pi)e^+\nu_e$~\cite{Ablikim:2015ixa,Ablikim:2015qgt,Ablikim:2017lks} 
\begin{gather}
	\mathcal{B}_{D^+\to K_S^0e^+\nu_e} = [8.604\pm0.056(\rm stat)\pm0.151(syst)] \%, \\ 
	\mathcal{B}_{D^+\to \pi^0e^+\nu_e} = [0.363\pm0.008(\rm stat)\pm0.005(syst)] \%,\\
		\mathcal{B}_{D^0\to K^-e^+\nu_e} = [3.505\pm0.014(\rm stat)\pm0.033(syst)] \%, \\
		 \mathcal{B}_{D^0\to \pi^-e^+\nu_e} = [0.295\pm0.004(\rm stat)\pm0.003(syst)] \%,\\
		 	\mathcal{B}_{D^+\to K_L^0e^+\nu_e} = [4.482\pm0.027(\rm stat)\pm0.103(syst)] \%,
\end{gather}
and form factors~\cite{Ablikim:2015ixa,Ablikim:2015qgt,Ablikim:2017lks} of $D\to K(\pi)e^+\nu_e$
\begin{gather}
f_+^K(0)[D^+\to K_S^0e^+\nu_e]=[0.7248\pm0.0041(\rm stat)\pm0.0115(syst)]\%,\\
f_+^K(0)[D^0\to K^-e^+\nu_e] = [0.7368\pm0.0026(\rm stat)\pm0.0036(syst)]\%,\\
f_+^\pi(0)[D^+\to \pi^0e^+\nu_e]=[0.6216\pm0.0115(\rm stat)\pm0.0035(syst)]\%,\\
f_+^\pi(0)[D^+\to \pi^0e^+\nu_e]=[0.6372\pm0.0080(\rm stat)\pm0.0044(syst)]\%,\\
f_+^K(0)[D^+\to K_L^0e^+\nu_e]=[0.748\pm0.007(\rm stat)\pm0.012(syst)]\%.
\end{gather}
Figure~\ref{projections_of_Fits_onto_ff_q2_stat_Kev}, \ref{projections_of_Fits_onto_ff_q2_stat_piev} and \ref{fig:FF} shows the projections of form factor on the fit to partial decay rates of $D\to K(\pi)e^+\nu_e$ except for $D^+\to K_L^0e^+\nu_e$.


\subsection{$D\to K^-(\pi) \mu^+\nu_\mu$}
\begin{figure}
\includegraphics[width=16cm]{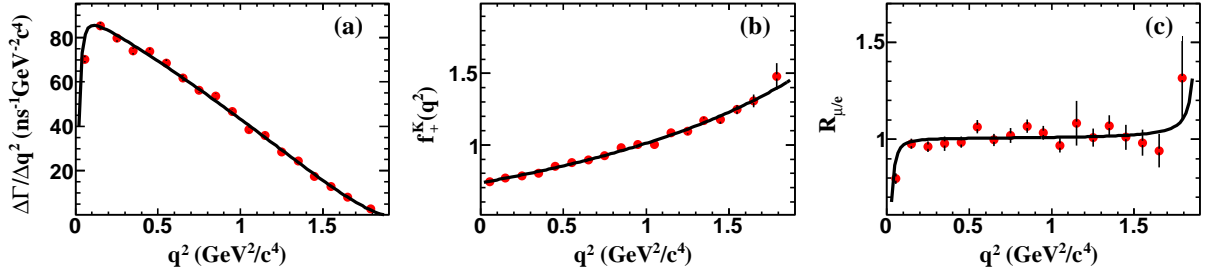}
\caption{The fit to the partial decay rates of $D^0\to K^-\mu^+\nu_\mu$ (up left), the projection
to the hadronic form factor (up right) and LFU test in various $q^2$ intervals (right).\label{fig:kmunu}}
\end{figure}

Muon channels also provide a chance to improve the precision of measurement on form factor $f_+^K(0)$, and more important, recent tension of LFU between $\tau^+$ and $\mu^+$~\cite{Lees:2012xj,Aaij:2015yra,Hirose:2017dxl} need improved understanding in charm sector.
Using 2.93 fb$^{-1}$ data at $\sqrt{s}=3.773$~GeV, the BF of $D^0\to K^-\mu^+\nu_\mu$ is measured to be $[3.413\pm0.019(\rm stat)\pm0.035(syst)]\%$. Combining with $\mathcal{B}_{D^0\to K^-e^+\nu_e}$, we have
\begin{equation}
	R_{K^-} = \frac{\Gamma(D^0\to K^-\mu^+\nu_\mu)}{\Gamma(D^0\to K^-e^+\nu_e)} = 0.974\pm0.007(\rm stat)\pm0.012(syst),
\end{equation}
where lifetime of $D^0$ is cancelled. With the same data and fitting method as previous electron channel, we obtain $f^K_+(0) = 0.7327\pm0.0039(\rm stat)\pm0.0030(\rm syst)$~\cite{Ablikim:2018evp}. Figure~\ref{fig:kmunu} shows the projection of form factor on the fit to partial decay rates. Combining with our previous measurement, LFU test is performed with
\begin{equation}
	R_{K^-}=\frac{\Gamma(D^0\to K^-\mu^+\nu_\mu)}{\Gamma(D^0\to K^-e^+\nu_e)}=0.974\pm0.007(\rm stat)\pm0.012(\rm syst).
\end{equation}
There is no deviation lager than 2$\sigma$ from 1 in $q^2$ interval (0.2,\,1.5)~GeV$^2$/c$^4$ as Fig~\ref{fig:kmunu} shows.  For the pion channel, the BF of $D\to\pi\mu^+\nu_\mu$~\cite{Ablikim:2018frk} is measured to be $\mathcal{B}_{D^0\to \pi^-\mu^+\nu_\mu}=[0.272\pm0.008(\rm stat)\pm0.006(\rm syst)]\%$ and $\mathcal{B}_{D^+\to \pi^-\mu^+\nu_\mu}=[0.350\pm0.011(\rm stat)\pm0.010(\rm syst)]\%$. Using these results along with $\mathcal{B}_{D\to \pi e^+\nu_e}$, we have
\begin{gather}
	R_{\pi^-}=\frac{\Gamma(D^0\to \pi^-\mu^+\nu_\mu)}{\Gamma(D^0\to \pi^-e^+\nu_e)}=0.922\pm0.030(\rm stat)\pm0.022(\rm syst),\\
	R_{\pi^0}= \frac{\Gamma(D^0\to \pi^0\mu^+\nu_\mu)}{\Gamma(D^0\to \pi^0e^+\nu_e)}=0.964\pm0.037(\rm stat)\pm0.026(\rm syst). 
\end{gather}
These results show no significant deviations from the standard model predictions.

\subsection{$D^+_s\to \eta^{(\prime)} e^+\nu_e$}
\begin{figure}
\centering
     \mbox{
    \begin{overpic}[width=16cm,angle=0]{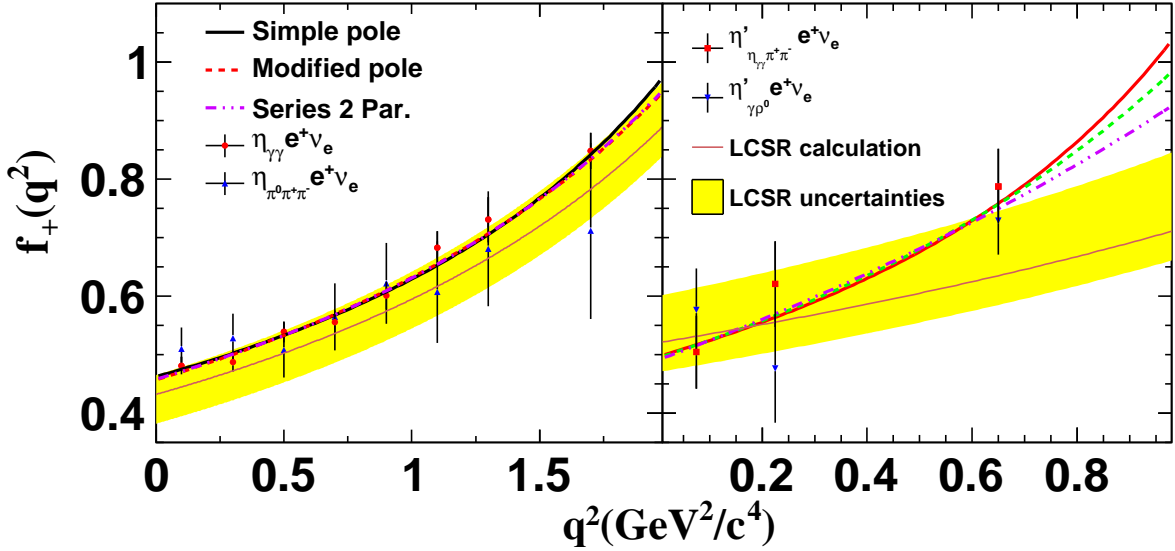}
    \end{overpic}
    }
\caption{\small Projections of the fits to partial decay rate of $D^+_s\to \eta^{(\prime)}e^+\nu_e$. Dots with error bars are data. Curves are the fits as described in text. Pink lines with yellow bands are the LCSR calculations with uncertainties. \label{fig:combine_FF}}
  
\end{figure}

BESIII measure the absolute BFs for semi-leptonic $D^+_s\to\eta^{(\prime)} e^+\nu_e$ decays~\cite{Ablikim:2019rjz} with improved precision. The preliminary results are $\mathcal{B}_{D^+_s\to\eta e^+\nu_e}=[2.32\pm0.06(\rm stat)\pm0.06(\rm syst)]\%$ and $\mathcal{B}_{D^+_s\to\eta e^+\nu_e}=[0.82\pm0.07(\rm stat)\pm0.03(\rm syst)]\%$ by a simultaneous fits on $\eta\to\gamma\gamma$ and $\eta\pi^+\pi^-\pi^0$ for $\eta$ mode and $\eta^\prime\to\eta_{\gamma\gamma}\pi^+\pi^-$ and $\eta^\prime\to\gamma\pi^+\pi^-$ for $\eta^\prime$ mode. Combing the our previous measurement on $\mathcal{B}_{D^+\to\eta^{(\prime)} e^+\nu_e}$~\cite{Ablikim:2018lfp}, the $\eta-\eta^\prime$ mixing angle is determined to be $\phi_P=(40.2\pm1.4(\rm stat)\pm0.5(\rm syst))^\circ$. And for the first time, the experimental measurement of the dynamics of  $D^+_s\to \eta^{(\prime)} e^+\nu_e$ are performed, the products of the hadronic form factor $f_+^{\eta^{(\prime)}}(0)$ and $|V_{cs}|$ are extracted with different form factor parameterizations. Figure~\ref{fig:combine_FF}  shows the projection of form factor on the fit to partial decay rates, where the yellow band comes from Light cone sum rule~\cite{Offen:2013nma}. For the two parameter series expansion, the preliminary results are $f_+^\eta(0)|V_{cs}| = 0.446\pm0.005(\rm stat)\pm0.004(\rm syst)$ and $f_+^{\eta^\prime}(0)|V_{cs}| = 0.477\pm0.049(\rm stat)\pm0.011(\rm syst)$. Taking $|V_{cs}|$ from the CKMfitter as input, we determine preliminary $f_+^\eta (0)= 0.458\pm0.005(\rm stat)\pm0.004(\rm syst)$ and $f_+^{\eta^\prime}(0) = 0.490\pm0.050(\rm stat)\pm0.011(\rm syst)$. Alternatively, using the $f_+^{\eta^{(\prime)}}(0)$  calculated by light-cone sum rules leads to $|V_{cs}| = 1.032\pm0.012(\rm stat)\pm0.009(\rm syst)\pm0.079(\rm theo)$ and $0.917\pm0.094(\rm stat)\pm0.021(\rm syst)\pm0.155(\rm theo)$, respectively.

\subsection{$D^+_s\to K^{0(*)} e^+\nu_e$}
Using the  data sample collected at $\sqrt{s} = 4.178$~GeV, BESIII measured $D^+_s\to K^{0(*)} e^+\nu_e$~\cite{Ablikim:2018upe}. The preliminary results are $\mathcal{B}_{D_s^+\to K^{0} e^+\nu_e}=[3.25\pm0.38(\rm stat)\pm0.16(\rm syst)]\%$ and $\mathcal{B}_{D^+_s\to K^{0*} e^+\nu_e}=[2.37\pm0.26(\rm stat)\pm0.20(\rm syst)]\%$. The first measurements of the hadronic form-factor parameters are obtained. The preliminary result for $D_s^+\to K^{0} e^+\nu_e$ is $f_+^{K}=0.720\pm0.084(\rm stat)\pm0.013(\rm syst)$, and for $D_s^+\to K^{0*} e^+\nu_e$,  the preliminary form-factor ratios are $r_V=V(0)/A_1(0) = 1.67\pm0.34(\rm stat)\pm0.016(\rm syst)$ and $r_2 = A_2(0)/A_1(0) = 0.77\pm0.28(\rm stat)\pm0.07(\rm syst)$.

\begin{figure}
\centering
\begin{minipage}[t]{0.49\linewidth}
\includegraphics[width=0.9\linewidth]{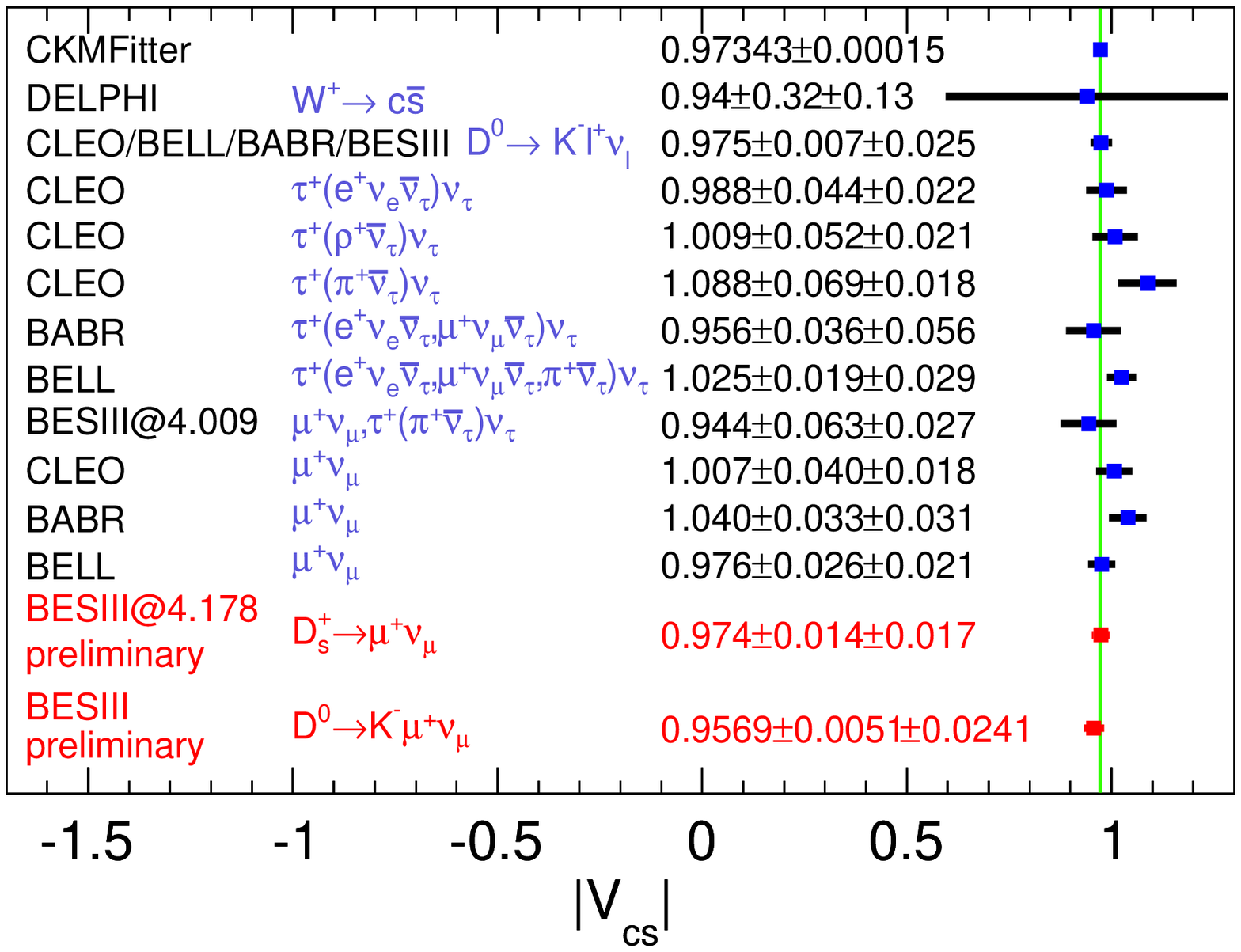}
\caption{\label{fig:Vcs} Comparison of $|V_{cs}|$.}
\end{minipage}
\begin{minipage}[t]{0.49\linewidth}
\includegraphics[width=0.9\linewidth]{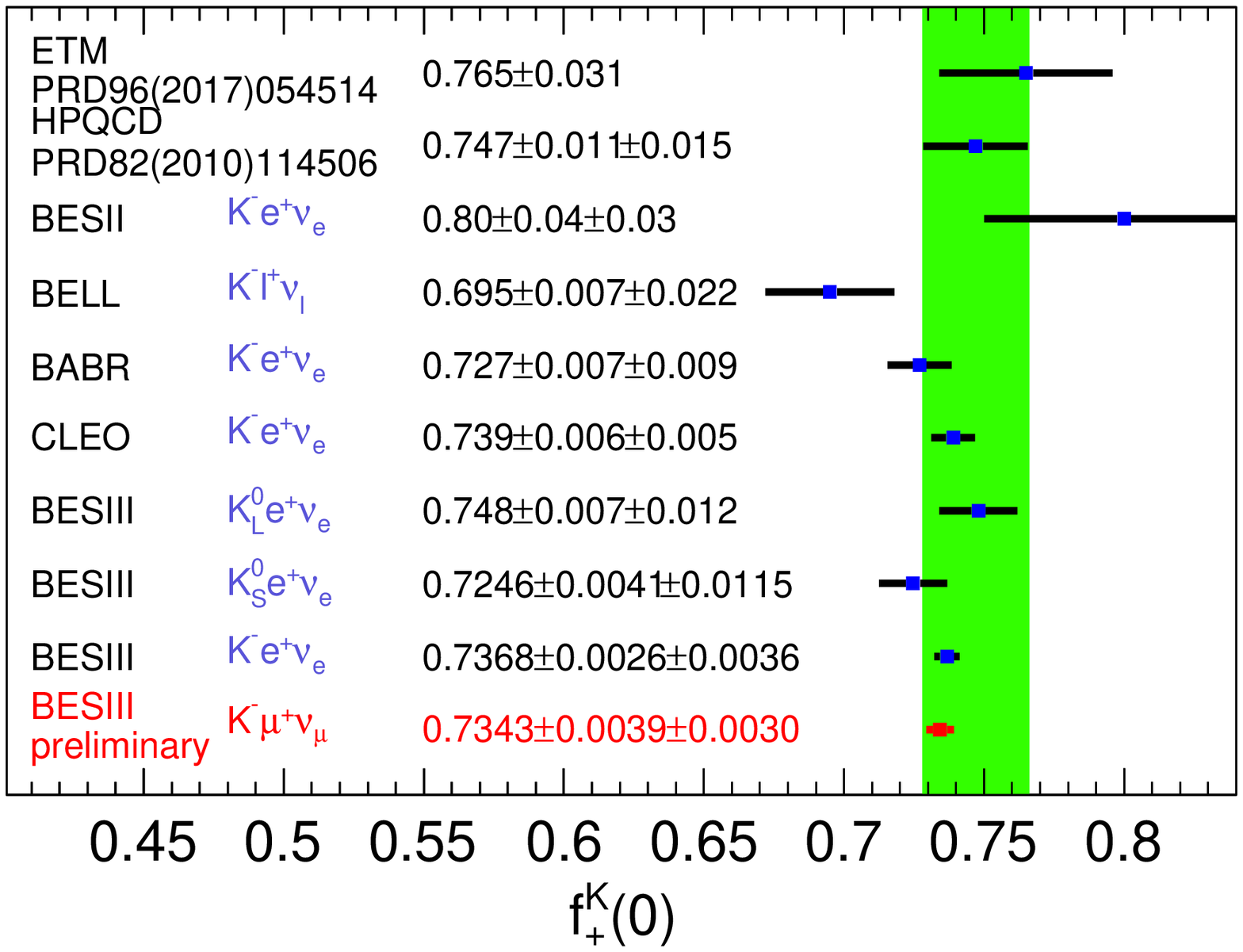}
\caption{\label{fig:fk} Comparison of $f^K_+(0)$.}
\end{minipage}
\end{figure}

\section{Summary}
In summary, with the word's largest $D\bar{D}$ samples near threshold, precision measurements of BFs of $D_{(s)}^+\to\ell^+\nu_\ell$, $D\to \bar K(\pi)\ell^+\nu_\ell$, $D_{(s)0}^+\to\eta^{\prime} e^+\nu_e$ and $D_s^+\to K^{0(*)}e^+\nu_e$ are performed at BESIII. In these decays, the form factor of $f^{D_s\to \eta}$, $f^{D_s\to K^{0(*)}}$ are extracted for the first time. Besides, CKM absolute matrix $|V_{cs(d)}|$, D meson decay constant $f_{D_s^+}$ and hadronic from factor $f_+^{D\to K}$ are also determined. 

  Meanwhile, LFU test using (semi-)leptonic $D$ decays is performed at BESIII, and no
significant deviation from the SM prediction is found at current statistics.

\section*{Funding information}
The author thanks for the support by Jonit Large-Scale Scientific
Facility Funds of the National Natural Science Foundation of China and the Chinese
Academy of Sciences under Contract No. U1532257.
%\section*{References}


\begin{thebibliography}{99}

  
 \bibitem{Ablikim:2015gyp} 
  M.~Ablikim {\it et al.} [BESIII Collaboration],
  %``Measurement of the form factors in the decay $D^+ \to \omega e^+ \nu_{e}$ and search for the decay $D^+ \to \phi e^+ \nu_{e}$,''
  Phys.\ Rev.\ D {\bf 92}, 071101 (2015).
 
 \bibitem{Ablikim:2015qgt} 
  M.~Ablikim {\it et al.} [BESIII Collaboration],
  %``Study of decay dynamics and $CP$ asymmetry in $D^+ \to K^0_L e^+ \nu_e$ decay,''
  Phys.\ Rev.\ D {\bf 92}, 112008 (2015).
 \bibitem{Ablikim:2015ixa} 
  M.~Ablikim {\it et al.} [BESIII Collaboration],
  %``Study of Dynamics of $D^0 \to K^- e^+ \nu_{e}$ and $D^0\to\pi^- e^+ \nu_{e}$ Decays,''
  Phys.\ Rev.\ D {\bf 92}, 072012 (2015).

 \bibitem{Ablikim:2015mjo} 
  M.~Ablikim {\it et al.} [BESIII Collaboration],
  %``Study of $D^{+} \to K^{-} \pi^+ e^+ \nu_e$,''
  Phys.\ Rev.\ D {\bf 94}, 032001 (2016).

 \bibitem{Ablikim:2016sqt} 
  M.~Ablikim {\it et al.} [BESIII Collaboration],
  %``Improved measurement of the absolute branching fraction of $D^{+}\rightarrow \bar{K}^0 \mu ^{+}\nu _{\mu }$,''
  Eur.\ Phys.\ J.\ C {\bf 76}, no. 7, 369 (2016).
 
  
  \bibitem{Ablikim:2017twd} 
  M.~Ablikim {\it et al.} [BESIII Collaboration],
  %``Search for the radiative leptonic decay $D^{+}\to \gamma e^{+} {\nu}_{e}$,''
  Phys.\ Rev.\ D {\bf 95}, no. 7, 071102 (2017).
 \bibitem{Ablikim:2017lks} 
  M.~Ablikim {\it et al.} (BESIII Collaboration),
  %``Analysis of $D^+\to\bar K^0e^+\nu_e$ and $D^+\to\pi^0e^+\nu_e$ semileptonic decays,''
  Phys.\ Rev.\ D {\bf 96}, 012002 (2017).

\bibitem{Ablikim:2017tdj} 
  M.~Ablikim {\it et al.} [BESIII Collaboration],
  %``Search for the rare decay $D^+ \to D^0 e^+\nu_e$,''
  Phys.\ Rev.\ D {\bf 96}, 092002 (2017).
  
 
  
 \bibitem{Ablikim:2017omq} 
  M.~Ablikim {\it et al.}[BESIII Collaboration],
  %``Measurements of the branching fractions for the semi-leptonic decays $D^+_s\to\phi e^{+}\nu_{e}$, $\phi \mu^{+}\nu_{\mu}$, $\eta \mu^{+}\nu_{\mu}$ and $\eta'\mu^{+}\nu_{\mu}$,''
  Phys.\ Rev.\ D {\bf 97}, 012006 (2018).


 \bibitem{Ablikim:2018evp} 
  M.~Ablikim {\it et al.} [BESIII Collaboration],
  %``Study of the $D^0\to K^-\mu^+\nu_\mu$ dynamics and test of lepton flavor universality with $D^0\to K^-\ell^+\nu_\ell$ decays,''
  arXiv:1810.03127 [hep-ex].
 
 
 
 \bibitem{Ablikim:2018lfp} 
  M.~Ablikim {\it et al.} [BESIII Collaboration],
  %``Study of the decays $D^+\rightarrow\eta^{(\prime)} e^+\nu_{e}$,''
  Phys.\ Rev.\ D {\bf 97}, 092009 (2018).
  %doi:10.1103/PhysRevD.97.092009
  %[arXiv:1803.05570 [hep-ex]].
  %%CITATION = doi:10.1103/PhysRevD.97.092009;%%
  %1 citations counted in INSPIRE as of 23 Nov 2018
  
  \bibitem{Ablikim:2018frk} 
  M.~Ablikim {\it et al.} [BESIII Collaboration],
  %``Measurement of the branching fraction for the semi-leptonic decay $D^{0(+)}\to \pi^{-(0)}\mu^+\nu_\mu$ and test of lepton universality,''
  Phys.\ Rev.\ Lett.\  {\bf 121}, 171803 (2018).
 
  \bibitem{Ablikim:2018ffp} 
  M.~Ablikim {\it et al.} [BESIII Collaboration],
  %``Observation of the Semileptonic Decay $D^0 \to a_0(980)^- e^+ \nu_e$ and Evidence for $D^+ \to a_0(980)^0 e^+ \nu_e$,''
  Phys.\ Rev.\ Lett.\  {\bf 121}, no. 8, 081802 (2018).
 
\bibitem{Ablikim:2018jun} 
  M.~Ablikim {\it et al.}[BESIII Collaboration],
  %``Determination of the pseudoscalar decay constant $f_{D_s^+}$ via $D_s^+\to\mu^+\nu_\mu$,''
  arXiv:1811.10890 [hep-ex].
     
\bibitem{Ablikim:2013uvu} 
  M.~Ablikim {\it et al.} [BESIII Collaboration],
  %``Precision measurements of $B(D^+ \rightarrow \mu^+ \nu_{\mu})$, the pseudoscalar decay constant $f_{D^+}$, and the quark mixing matrix element $|V_{\rm cd}|$,''
  Phys.\ Rev.\ D {\bf 89}, 051104 (2014).
  %doi:10.1103/PhysRevD.89.051104
  %[arXiv:1312.0374 [hep-ex]].
  %%CITATION = doi:10.1103/PhysRevD.89.051104;%%
  %40 citations counted in INSPIRE as of 12 Nov 2018
 %\bibitem{Patrignani:2016xqp} M.~Tanabashi {\it et al.} (Particle Data Group), Phys.\ Rev. D {\bf 98}, 030001 (2018).
  
  
  \bibitem{Ablikim:2016rqq} 
  M.~Ablikim {\it et al.} [BESIII Collaboration],
  %``Measurements of the absolute branching fractions for $D_{s}^{+}\rightarrow\eta e^{+}\nu_{e}$ and $D_{s}^{+}\rightarrow\eta^{\prime} e^{+}\nu_{e}$,''
  Phys.\ Rev.\ D {\bf 94}, 112003 (2016)
  
  \bibitem{Ablikim:2016duz} 
  M.~Ablikim {\it et al.} [BESIII Collaboration],
  %``Measurement of the $D_s^+ \to \ell^+\nu_\ell$ branching fractions and the decay constant $f_{D_s^+}$,''
  Phys.\ Rev.\ D {\bf 94}, 072004 (2016).
  
  \bibitem{Ablikim:2018upe} 
  M.~Ablikim {\it et al.} [BESIII Collaboration],
  %``First measurement of the form factors in $D^+_{s}\rightarrow K^0 e^+\nu_e$ and $D^+_{s}\rightarrow K^{*0} e^+\nu_e$ decays,''
  arXiv:1811.02911 [hep-ex].
  
  \bibitem{Ablikim:2019rjz} 
  M.~Ablikim {\it et al.} [BESIII Collaboration],
  %``Measurement of the dynamics of the decays ${ D_s^+ \rightarrow \eta^{(\prime)} e^{+} \nu_e}$,''
  arXiv:1901.02133 [hep-ex].
  
  \bibitem{Ablikim:2015prg} 
  M.~Ablikim {\it et al.} [BESIII Collaboration],
  %``Measurement of the absolute branching fraction for $\Lambda^+_{c}\to \Lambda e^+\nu_e$,''
  Phys.\ Rev.\ Lett.\  {\bf 115}, 221805 (2015).  
  
  
  \bibitem{Ablikim:2018woi} 
  M.~Ablikim {\it et al.} [BESIII Collaboration],
  %``Measurement of the absolute branching fraction of the inclusive semileptonic $\Lambda_c^+$ decay,''
  arXiv:1805.09060 [hep-ex].
  
  \bibitem{Lubicz:2017syv} 
  V.~Lubicz {\it et al.} [ETM Collaboration],
  %``Scalar and vector form factors of $D \to \pi(K) \ell \nu$ decays with $N_f=2+1+1$ twisted fermions,''
  Phys.\ Rev.\ D {\bf 96}, no. 5, 054514 (2017).
  
  \bibitem{Riggio:2017zwh} 
  L.~Riggio, G.~Salerno and S.~Simula,
  %``Extraction of $|V_{cd}|$ and $|V_{cs}|$ from experimental decay rates using lattice QCD $D \to \pi(K) \ell \nu$ form factors,''
  Eur.\ Phys.\ J.\ C {\bf 78}, no. 6, 501 (2018).
  %doi:10.1140/epjc/s10052-018-5943-5
  %[arXiv:1706.03657 [hep-lat]].
\bibitem{Aubin:2004ej} 
  C.~Aubin {\it et al.} [Fermilab Lattice and MILC and HPQCD Collaborations],
  %``Semileptonic decays of D mesons in three-flavor lattice QCD,''
  Phys.\ Rev.\ Lett.\  {\bf 94}, 011601 (2005).
  %doi:10.1103/PhysRevLett.94.011601
  %[hep-ph/0408306].
  \bibitem{Ball:1991bs} 
  P.~Ball, V.~M.~Braun and H.~G.~Dosch,
  %``Form-factors of semileptonic D decays from QCD sum rules,''
  Phys.\ Rev.\ D {\bf 44}, 3567 (1991).
  
  \bibitem{Na:2010uf} 
  H.~Na, C.~T.~H.~Davies, E.~Follana, G.~P.~Lepage and J.~Shigemitsu,
  %``The $D \rightarrow K, l \nu$ Semileptonic Decay Scalar Form Factor and $|V_{cs}|$ from Lattice QCD,''
  Phys.\ Rev.\ D {\bf 82}, 114506 (2010).
  
  \bibitem{Bazavov:2017lyh} 
  A.~Bazavov {\it et al.},
  %``$B$- and $D$-meson leptonic decay constants from four-flavor lattice QCD,''
  Phys.\ Rev.\ D {\bf 98}, 074512 (2018).
  
  \bibitem{Bazavov:2014wgs} 
  A.~Bazavov {\it et al.} [Fermilab Lattice and MILC Collaborations],
  %``Charmed and light pseudoscalar meson decay constants from four-flavor lattice QCD with physical light quarks,''
  Phys.\ Rev.\ D {\bf 90}, 074509 (2014).
  
  \bibitem{Boyle:2017jwu} 
  P.~A.~Boyle, L.~Del Debbio, A.~Jüttner, A.~Khamseh, F.~Sanfilippo and J.~T.~Tsang,
  %``The decay constants ${\mathbf{f_D}}$ and ${\mathbf{f_{D_{s}}}}$ in the continuum limit of ${\mathbf{N_f=2+1}}$ domain wall lattice QCD,''
  JHEP {\bf 1712}, 008 (2017).

  \bibitem{Yang:2014sea} 
  Y.~B.~Yang {\it et al.},
  %``Charm and strange quark masses and $f_{D_s}$ from overlap fermions,''
  Phys.\ Rev.\ D {\bf 92}, 034517 (2015).
  
  \bibitem{Bazavov:2011aa} 
  A.~Bazavov {\it et al.} [Fermilab Lattice and MILC Collaborations],
  %``B- and D-meson decay constants from three-flavor lattice QCD,''
  Phys.\ Rev.\ D {\bf 85}, 114506 (2012).
  
  \bibitem{Hwang:2009qz} 
  C.~W.~Hwang,
  %``SU(3) symmetry breaking in decay constants and electromagnetic properties of pseudoscalar heavy mesons,''
  Phys.\ Rev.\ D {\bf 81}, 054022 (2010).
  
  \bibitem{Becirevic:1998ua} 
  D.~Becirevic, P.~Boucaud, J.~P.~Leroy, V.~Lubicz, G.~Martinelli, F.~Mescia and F.~Rapuano,
  %``Nonperturbatively improved heavy - light mesons: Masses and decay constants,''
  Phys.\ Rev.\ D {\bf 60}, 074501 (1999).
  
  \bibitem{Na:2012iu} 
  H.~Na, C.~T.~H.~Davies, E.~Follana, G.~P.~Lepage and J.~Shigemitsu,
  %``$|V_{cd}|$ from D Meson Leptonic Decays,''
  Phys.\ Rev.\ D {\bf 86}, 054510 (2012).
  
  \bibitem{Aubin:2005ar} 
  C.~Aubin {\it et al.},
  %``Charmed meson decay constants in three-flavor lattice QCD,''
  Phys.\ Rev.\ Lett.\  {\bf 95}, 122002 (2005).
  
  \bibitem{Follana:2007uv} 
  E.~Follana {\it et al.} [HPQCD and UKQCD Collaborations],
  %``High Precision determination of the pi, K, D and D(s) decay constants from lattice QCD,''
  Phys.\ Rev.\ Lett.\  {\bf 100}, 062002 (2008).
  
  \bibitem{Dimopoulos:2011gx} 
  P.~Dimopoulos {\it et al.} [ETM Collaboration],
  %``Lattice QCD determination of m_b, f_B and f_Bs with twisted mass Wilson fermions,''
  JHEP {\bf 1201}, 046 (2012).
  
  \bibitem{Chiu:2005ue} 
  T.~W.~Chiu, T.~H.~Hsieh, J.~Y.~Lee, P.~H.~Liu and H.~J.~Chang,
  %``Pseudoscalar decay constants f(D) and f(D(s)) in lattice QCD with exact chiral symmetry,''
  Phys.\ Lett.\ B {\bf 624}, 31 (2005).
  
  \bibitem{Lellouch:2000tw} 
  L.~Lellouch {\it et al.} [UKQCD Collaboration],
  %``Standard model matrix elements for neutral B meson mixing and associated decay constants,''
  Phys.\ Rev.\ D {\bf 64}, 094501 (2001).
  
  \bibitem{Badalian:2007km} 
  A.~M.~Badalian, B.~L.~G.~Bakker and Y.~A.~Simonov,
  %``Decay constants of the heavy-light mesons from the field correlator method,''
  Phys.\ Rev.\ D {\bf 75}, 116001 (2007).
  
  
  
  
  
  \bibitem{Silverman:1988gc} 
  D.~Silverman and H.~Yao,
  %``Relativistic Treatment of Light Quarks in $D$ and $B$ Mesons and $W$ Exchange Weak Decays,''
  Phys.\ Rev.\ D {\bf 38}, 214 (1988).

  
\bibitem{Lees:2012xj} 
  J.~P.~Lees {\it et al.} [BaBar Collaboration],
  %``Evidence for an excess of $\bar{B} \to D^{(*)} \tau^-\bar{\nu}_\tau$ decays,''
  Phys.\ Rev.\ Lett.\  {\bf 109}, 101802 (2012).
 
 \bibitem{Aaij:2015yra} 
  R.~Aaij {\it et al.} [LHCb Collaboration],
  %``Measurement of the ratio of branching fractions $\mathcal{B}(\bar{B}^0 \to D^{*+}\tau^{-}\bar{\nu}_{\tau})/\mathcal{B}(\bar{B}^0 \to D^{*+}\mu^{-}\bar{\nu}_{\mu})$,''
  Phys.\ Rev.\ Lett.\  {\bf 115}, 111803 (2015)
  Erratum: [Phys.\ Rev.\ Lett.\  {\bf 115}, 159901 (2015)].
  
 \bibitem{Hirose:2017dxl} 
  S.~Hirose {\it et al.} [Belle Collaboration],
  %``Measurement of the $\tau$ lepton polarization and $R(D^*)$ in the decay $\bar{B} \rightarrow D^* \tau^- \bar{\nu}_\tau$ with one-prong hadronic $\tau$ decays at Belle,''
    Phys.\ Rev.\ D {\bf 97}, 012004 (2018).

 
  
  \bibitem{Offen:2013nma}
  N.~Offen, F.~A.~Porkert and A.~Sch{\"a}fer,
  Phys.\ Rev.\ D {\bf 88}, 034023 (2013).





\end{thebibliography}
\end{document}